\documentclass[12pt]{elsarticlealt}

\usepackage{graphicx}
\usepackage{dcolumn}
\usepackage{bm}

\def\ii{\'{\i }}

\def\infinito{\smash{\mathop{\longrightarrow}\limits_{\raise3pt\hbox{$_{y_b \to \infty}$}}}}

\def\doisdeltas{\smash{\mathop{\longrightarrow}\limits_{\raise3pt\hbox{$_{\Delta >\Delta_0 (y_b)}$}}}}

\usepackage{epsfig}

\def\menorsim{\smash{\mathop{<}\limits_{\raise3pt\hbox{$\sim$}}}}

\def\pigrande{\displaystyle\prod_{i=1}^{\nu_{ _N}}}

\begin{document}

\begin{frontmatter}

\title{Long range forward-backward rapiditiy correlations in proton-proton collisions at LHC}%

\author{P. Brogueira}
\ead{pedro@fisica.ist.utl.pt}
\address{Departamento de F\ii sica, IST, Av. Rovisco Pais, 1049-001 Lisboa, Portugal }
\author{J. Dias de Deus}
\ead{jdd@fisica.ist.utl.pt}
\address{CENTRA, Departamento de F\ii sica, IST, Av. Rovisco Pais, 1049-001 Lisboa, Portugal}
\author{C. Pajares}
\ead{pajares@fpaxp1.usc.es}
\address{IGFAE and Departamento de Fisica de Particulas, Univ. of Santigo de Compostela, 15706, Santiago de Compostela, Spain}

\date{\today}%

\begin{abstract}
We argue that string percolation is in the origin of i) an approximately flat rapidity distribution and of ii) an approximately constant forward-backward correlation parameter $b$ over a substantial fraction of the available rapidity. Predictions are given for $pp$ collisions at LHC, $\sqrt{s}=14$TeV and $\sqrt{s}=5.5$TeV.
\end{abstract}


\end{frontmatter}

In $AA$ (nucleus-nucleus or hadron-hadron) high energy collisions one naturally expects positive Forward-Backward (F-B) correlations to occur. This happens because the overlap area of interaction acts both in the forward and in the backward direction. If the impact parameter is small, many particles are emitted forward, and many particles are emitted backward. On the contrary, if the impact parameter is large, few particles are emitted forward and few particles are emitted backward. Note that in addition to these F-B correlations, local correlations in rapidity exist due, for instance, to resonance decays. As these short-range correlations are not supposed to depend on impact parameter, it is clear that F-B correlations increasingly dominate with the decrease of impact parameter, or the increase of the number of participants $N_{part.}$. (See, for instance [1]).

The observation of F-B correlations is then not a surprise. What was not anticipated was that the strength of the F-B correlations was practically constant over a large rapidity interval, perhaps of the order of the beam rapidity interval $\Delta Y$. Recent results at RHIC (including the discovery of the ridge phenomenon) show that the F-B correlations extend over a large region in rapidity [2,3,4,5].

Monte Carlo event generators, such as HIJING, predict a fast decrease of the F-B correlation with the rapidity (pseudorapidity) interval $\Delta \eta$ [6]. The Parton String Model [7], which includes pair of strings fusion, shows a slower drecrease of the correlation with $\Delta \eta$, thus being in closer agreement with data (see [8]).

Models that can explain, in a straight forward manner, long range F-B correlations are models that introduce extended objects in rapidity as colour flux tubes [9] or strings [10]. 

In conventional F-B correlation studies $b$ is the F-B correlation parameter defined by the relation [11]
$$
\langle n_B\rangle \left( n_F \right) \sim b n_F \ ,\eqno(1)
$$
where $n_F$ is the number of particles in the forward rapidity window and $\langle n_B\rangle$ the average number of particles in the backward window. The correlation parameter $b$ is given by
$$
b\equiv \left[\langle n_F n_B\rangle - \langle n_F\rangle \langle n_B\rangle \right] / \left[ \langle n^2_F\rangle -\langle n_F\rangle^2 \right] \ . \eqno(2)
$$

Adopting the two step scenario [12,13], where extended objects are formed first, followed by local emissions of particles, assumed Poissonian, one obtains
$$
b = {\langle n_B\rangle / \langle n_F\rangle \over 1+ K / \langle n_F \rangle} \ , \eqno(3)
$$
Where $K$ is the inverse of the normalized variance of the number of collisions $\nu$ distribution:
$$
1/K = {\langle \nu^2 \rangle - \langle \nu \rangle^2 \over \langle \nu \rangle^2}\eqno(4)
$$
For symmetric windows $\langle n_B\rangle = \langle n_F\rangle$ and (3) becomes
$$
b= {1\over 1+ K / \langle n_F\rangle} \  . \eqno(5)
$$

It is now clear that in this formalism an almost $\Delta \eta$ independent F-B correlation means that the particle density distribution is approximately flat around mid rapidity. Particle densities and F-B correlations are a kind of direct image of the extended objects stretched between the interacting hadronic sheets [9].

In the dual string model (see, for instance, [1]) strings are constructed from partons, one from each nucleus or hadron, with Feynman $x$ values $x_+$ and $x_-$, respectively. At high energy or density most of the partons are wee partons (gluons) with, in general, $x_+ \simeq x_- \simeq 1/\sqrt{s}$, where $\sqrt{s}$ is the centre of mass energy. In such situation the formed strings are small, with fixed length $\Delta y_1$, in rapidity. One does expect the particle distribution to be peaked around mid rapidity, and being not flat at all in a large rapidity interval. But the region of $x_+ \simeq x_- \simeq 0$ is the region of saturation and percolation of strings. If $\langle N_s \rangle$ strings percolate the size in rapidity of the percolated strings is [14,15],
$$
\Delta y_{\langle N_s \rangle} = \Delta y_1 + 2 \ln \langle N_s \rangle \ . \eqno(6)
$$
In the  naive model with just these percolating strings, if we impose energy conservation [15] we obtain
$$
\langle N_s \rangle \sim s^{\lambda} \ , \eqno(7)
$$
with $\lambda =2/7$, such that
$$
\Delta y_{\langle N_s \rangle} =2\lambda \Delta Y\simeq {1\over 2}\Delta Y \ . \eqno(8)
$$
This simple model tells us that half of the full rapidity interval $\Delta Y$ is occupied by a flat distribution.

One can be more sophisticated and use an improved version of the step function, the Fermi-like distribution [16, 17] that fits RHIC/PHOBOS data and can be obtained by an evolution equation for the particle density [18], 
$$
{2\over N_{part.}} {dn\over dy} = {e^{\lambda Y}\over e^{\eta -(1-\alpha ) Y \over \delta}+1} \  ,  \  (\eta \geq 0) \  , \eqno(9)
$$   
where $N_{part.}$ is the number of participating nucleons, $Y$ is the beam rapidity and $\lambda , \alpha$ and $\delta$ are parameters (from fit to RHIC data: $\lambda =0.26$, $\alpha  =0.31$ and $\delta = 0.75$). The region of approximately flat particle density is between the inflexion point, at $\eta > 0$, and the corresponding point at $\eta < 0$:
$$
\Delta \eta_{Flat} \sim (1-\alpha) \Delta Y \simeq 0.7 \Delta Y \ . \eqno(10)
$$

Eq. (9), however,  strictly  speaking, is only valid for large values of $N_{part.}$. Having in mind, on one hand, the presence of nucleon-nucleon contribution and, on the other, multiple nucleon contributions, we write, in the high energy limit, see [19], 
$$
{2\over N_{part.}} {dn\over dy} = {1 \over 2} ( 1+ ( 1 - ( {2 \over N_{part.}})  ^{ 1 \over 3} ) ){e^{\lambda Y}\over e^{\eta -(1-\alpha ) Y \over \delta}+1} \   \  , \eqno(11)
$$ 
For large values of $N_{part.}$ we recover (9). For mid rapidity the validity of the low energy version of (11) was tested in [20].

Recently, in [19], the two step scenario was successfully applied to central {\it Au-Au} collisions, RHIC data at $\sqrt{s} =200 GeV$ [21]. In particular the correlation parameter $b$ was predicted to be fairly constant up to $\Delta \eta  \menorsim 3$. Predictions were also presented for $Pb -Pb$ at LHC, with an approximately constant $b$ extending over a region $\Delta \eta  \menorsim 6$.

Recent data on correlations for $pp$ collisions at RHIC [5] have shown that the rapidity correlation is peaked at $\eta =0$ -- which is typical of a non-percolating regime - and that no azimutal elliptic flow effects, contrary to central {\it Au-Au}, were detected. However there are strong indications that the physics of $pp$ collisions at very high energy is the same as the physics of heavy ion collisions at lower energy [22], and we then expect that the formalism of [19] can be applied to $pp$ collisions at LHC, $\sqrt{s}=14$TeV.

In fact, in percolation theory the transition to the percolation regime occurs for a critical value of the transverse density, 
$$
\eta_{AA} \equiv \left( {r \over R} \right)^2 \langle N_s \rangle \   \   \eqno(12)
$$ 
where $r$ is the single string radius, and $R$, the effective radius of the overlap region of interaction. Simple geometrical or Glauber like arguments give $R=R_p({N_{part.}\over 2})^{1 \over 3}$ and $\langle N_s\rangle \simeq \langle N_s^p\rangle({N_{part.} \over 2})^{4 \over 3}$, where $\langle N_s\rangle$ is the average number of strings and $\langle N_s^p\rangle$ the average number of strings in $pp$ collisions. In [10] it was estimated that in central $PbPb$ collisions the transitions with $\eta^c_{PbPb} \simeq 1.15$ occurred at SPS for $\sqrt{s} \simeq 20$GeV. By assuming that $\langle N_s^p\rangle \sim s^{2 \over 7}$ we estimate $\eta^c_{pp}$ to occur at LHC energies, for $\sqrt{s}_{pp} \simeq 6$TeV.

In order to estimate $b$, see (5), one needs $\langle n_F \rangle$ and $K$. The window multiplicity is estimated by using (11), under the assumption of a window with width $\delta \eta =0.2$ (as in [19]). The problem is to estimate $K$, (4), (in [19] $K$ was a free parameter).

\begin{figure}[t]
\begin{center}
\hskip -0.5 true cm
\includegraphics[width= 9 cm]{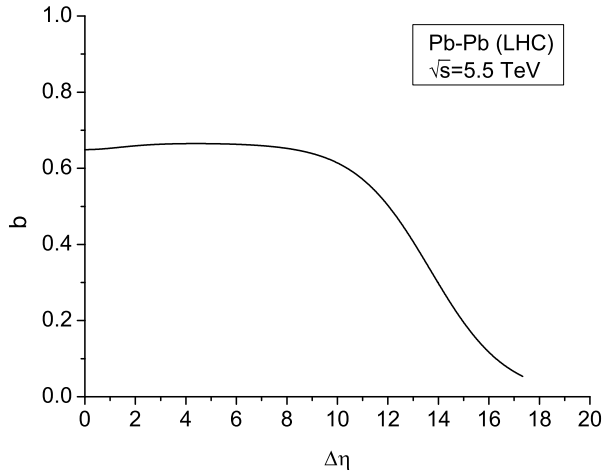}      
\end{center}
\caption{The F-B correlation parameter $b$ at LHC, $\sqrt{s}= 5.5$TeV, in $PbPb$ central collisions.}
\end{figure}

In a simple model - again a two step scenario - where nucleon-nucleon collisions are followed, in an independent manner, by parton-parton collisions, we can write [23],
$$
\varphi (\nu) = \Sigma \psi (\nu_N ) \pigrande p (\mu_i) \delta (\nu -\Sigma \mu_i )\ , \eqno(13)
$$
where $\varphi (\nu )$ is the probability of having $\nu$ collisions, $\psi (\nu_N )$ the probability of having $\nu_N$ nucleon-nucleon collisions and $p(\mu )$ the probability of having, in a nucleon-nucleon collision (say, $pp$), $\mu$ partonic collisions. From (13) we conclude,
$$
\langle \nu \rangle = \langle \nu_N \rangle \langle \mu \rangle \ , \eqno(14)
$$
and
\begin{figure}[t]
\begin{center}
\hskip -0.5 true cm
\includegraphics[width= 9 cm]{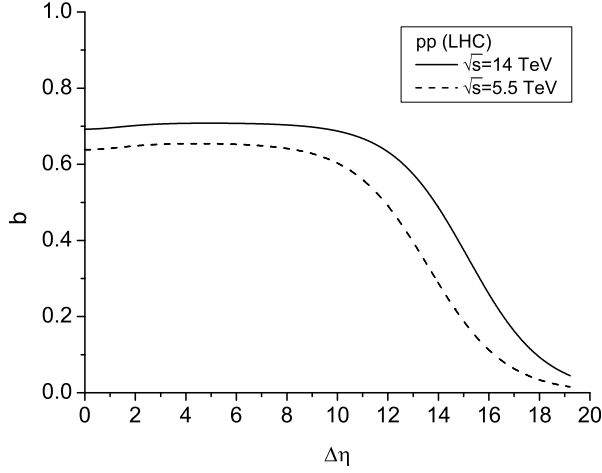}      
\end{center}
\caption{The F-B correlation parameter $b$ at LHC, $\sqrt{s}= 14$TeV, full line, and $\sqrt{s}= 5.5$TeV, dashed line, in $pp$ collisions.}
\end{figure}
$$
{\langle \nu^2 \rangle -\langle \nu \rangle^2 \over \langle \nu \rangle^2} = {\langle \nu_N^2 \rangle -\langle \nu_N \rangle^2 \over \langle \nu_N \rangle^2} + {\langle \mu^2 \rangle -\langle \mu \rangle^2 \over \langle \mu \rangle^2} {1\over \langle \nu_N \rangle} \ , \eqno(15)
$$
or,
$$
{1\over K}={1\over K_N} + {1\over k_{\mu}} {1\over \langle \nu_N \rangle} \ , \eqno(16)
$$
where $k_{\mu} \equiv \langle \mu \rangle^2/ (\langle \mu^2 \rangle -\langle \mu \rangle^2 )$, and $K_N \equiv \langle \nu_N \rangle^2 /(\langle \nu_N^2 \rangle -\langle \nu_N \rangle^2 )$. It is clear, from (16), that $K$ (for the overall collision distribution) is smaller than $K_N$ (for the nucleon collision distribution). For $N=1$ ($pp$ collisions) $K=k_{\mu}$. As $\langle n_F\rangle \sim dn/dy \sim {N_{part.} \over 2}$, (9), the correlation parameter $b$ increases with the $N_{part.}$, we have, in addition to (16), the relation (see (5)), 
$$
{K\over N_{part.}} < k_{\mu} \ . \eqno(17)
$$
Note that the quantities $K_N$ and $\langle \nu_N \rangle$ are fixed by Glauber calculus [24]: $K_N \simeq 300$ and $\langle \nu_N \rangle \simeq 800$.

\begin{figure}[t]
\begin{center}
\hskip -0.5 true cm
\includegraphics[width= 9 cm]{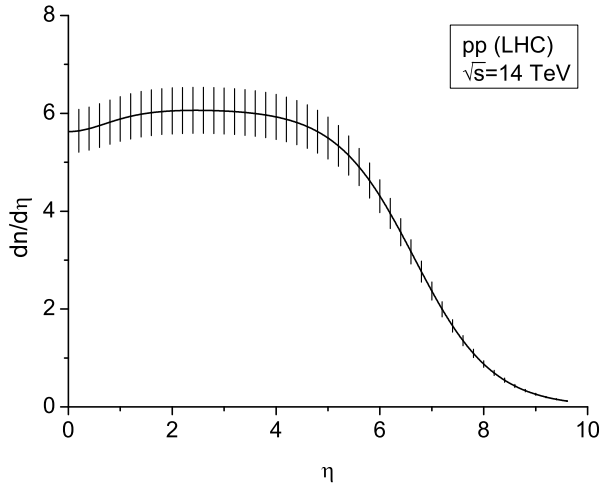}      
\end{center}
\caption{Particle density, $dn/d\eta$ in $pp$ collisions at $\sqrt{s}= 14$TeV. Errors due to a 3\% error atributed to the parameter $\lambda$.}
\end{figure}

By using (16) and (17) we are now in conditions of finding the parameters at LHC energies: $K\simeq 170$ for central $PbPb$ and $k_{\mu} \simeq 0.5$ for $pp$ collisions. The results for $b$ as a function of $\Delta \eta$ are shown in Figs. 1 and 2. In Fig. 3 we present our prediction for $dn/dy$ for $pp$ collisions at $\sqrt{s} = 14 TeV$. In Fig. 4, we show our expected dependence of ${2 / N_{Part.}}  \times {dn / d \eta} $ on $N_{Part.}$ (11) at LHC, $\sqrt{s}= 5.5$TeV.
\begin{figure}[t]
\begin{center}
\hskip -0.5 true cm
\includegraphics[width= 9 cm]{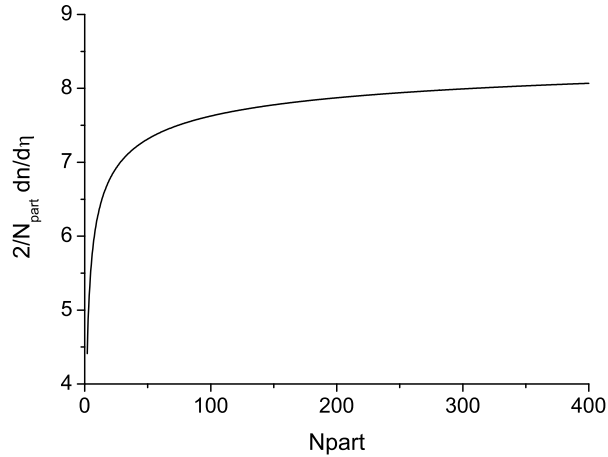}      
\end{center}
\caption{${2 / N_{Part.}}  \times {dn / d \eta} $ versus $N_{Part.}$ at LHC, $\sqrt{s}= 5.5$TeV.}
\end{figure}

We would like to finish with two remarks: i) Even in $pp$ collisions is possible to test the impact parameter dependence of the correlation parameter $b$, by comparing events with  different multiplicities (bellow the average and above the average, for instance). We expect higher multiplicity events to correspond to higher values for $b$. ii) In [25] it was made a prediction for $dn/d\eta (pp, \sqrt{s}= 14 {\rm TeV}) 70 \pm 8$ charged particles. We expect (see Fig.3) $80 \pm 7$ slightly larger than the value obtained in [25]. 

The possibility of existence of collective effects in $pp$ collisions at very high energy was considered in [26].

\bigskip

{\it Acknowledgments}

\bigskip
CP was supported by the Spanish Consolider-Ingenio 2010 programme CPAN, the project FPA2008-01177 and Conselleria de Educacion da Xunta de Galicia. JDD was supported by the FCT/Portugal project PPCDT/FIS/57568/2004.

\bigskip
\bigskip

References

\begin{enumerate}
\item A. Capella, U. Sukhatme, C.I. Tan and J. Tran Tanh Van, Phys. Rev. 236, 225 (1994).
\item J. Putschke, J. Phys. G34, 9679 (2007) [arXiv: nucl-ex/0701074].
\item J. Adams et al. [STAR Collabotation], J. Phys. G32, L37 (2006); Phys. Rev. C73, 064907 (2006); Phys. Rev. C72, 044902 (2005); D. Adamova et al. [CERES Collaboration], arXiv: 08032407 [nucl-ex]. 
\item S.J. Lindenbaun, R.S. Longacre and M. Kramer, Eur. Phys. J. C. 30, 241 (2003).
\item M. Daugherity [STAR Collaboration], arXiv: 08062121 [nucl-ex]; P. Sorensen, arxiv:0811.2959[nucl-ex].
\item X.N. Wang and M. Gyulassy, Phys. Rev. D44, 3501 (1991).
\item N:S. Amelin, M.A. Braun and C. Pajares, Phys. Lett. B306, 212 (1993).
\item B. Srivastava et al., Int. J. Mod. Phys., E16, 3371 (2008).
\item Larry McLerran, arXiv: 08074095 v1 [hep-ph] (2008); N. Armesto, L. McLerran, C. Pajares, Nucl. Phys. A781, 201 (2007); N. Armesto, M.A. Braun, C. Pajares, Phys. Rev. C75, 054902 (2007); S. Gavin, L. McLerran, G. Merchelli, arXiv: 0806.4718 v1 [nucl-th]; A. Dimitriu, F. Gelis, L. McLerran and R. Venugopalan, arXiv: 0804.3858 v1 [hep-ph] 2008.
\item N.S. Amelin, N. Armesto, M.A. Braun, E.G. Ferreiro and C. Pajares, Phys. Rev. Lett. 73, 2813 (1994); N. Armesto, M.A. Braun, E.G. Ferreiro and C. Pajares, Phys. Rev. Lett. 77, 3736 (1996).
\item A. Capella and A. Krzywicki, Phys. Rev. D18, 4120 (1978).
\item M.A. Braun, C. Pajares and V.V. Vecherin, Phys. Lett. B493, 54 (2000).
\item P. Brogueira and J. Dias de Deus, Phys. Lett. B 653, 202-205 (2007).
\item M.A. Braun, E.G. Ferreiro, F. del Moral and C. Pajares, Eur. Phys. J. C25, 249 (2002).
\item J. Dias de Deus, M.C. Espírito Santo, M. Pimenta and C. Pajares, Phys. Rev. Lett. 96, 162001 (2006).
\item J. Adams et al. [STAR Collaboration],nucl-ex/0511026.
\item P. Brogueira, J. Dias de Deus and C. Pajares, Phys. Rev. C75, 054908 (2007).
\item J. Dias de Deus and J.G. Milhano, Nucl. Phys. A795, 98 (2007) [arXiv:hep-ph/0701215].
\item P. Brogueira, J. Dias de Deus and J.G. Milhano, Phys. Rev. C76:064901, 2007. 
\item J. Dias de Deus and R. Ugoccioni, Phys. Lett. B 494, 53 (2000).
\item T.J. Tarnowsky [STAR Collaboration], nucl-ex/0606018; B.K. Srivastava [STAR Collaboration] nucl-ex/0702054.
\item L. Cunqueiro, J. Dias de Deus and C. Pajares, arXiv:0806.0523.
\item J. Dias de Deus, C. Pajares and C. Salgado, Phys. Lett. B407, 335 (1997).
\item V.V. Vechernin, hep-ph/0702141 (2007.
\item W. Busza, "Heavy Ion Collisions at the LHC - Last call for Predictions"', [arXiv-0711.0974], J. Phys. G. Nucl. Part. Phys. 35, 054001 (2008).
\item G. Feofilov, "Long-Range Multiplicity Correlation in $pp$-collisions as manifestation of collectivity effects in PYTHA: comparison to experimental data and predictions for ALICE", ALICE meetting, 31-03-2008.
\end{enumerate}
\end{document}